\documentclass[aps,prmaterials,reprint,amsmath,amssymb,superscriptaddress]{revtex4-2}

\usepackage{graphicx}
\usepackage{booktabs}
\usepackage{siunitx}
\usepackage{bm}
\usepackage{pgfplots}
\usepackage{pgfplotstable}
\usepackage{hyperref}
\hypersetup{hidelinks}
\pgfplotsset{compat=1.18}
\newcommand{\edot}{\dot{\epsilon}}
\newcommand{\ep}{\epsilon_{\rm p}}
\newcommand{\Eeff}{E_{\rm eff}}
\newcommand{\mus}{\mu_{\sigma}}
\newcommand{\vb}{v_{\rm b}}
\newcommand{\Deps}{\Delta\epsilon_{\rm b}}

\begin{document}

\title{From dynamic strain aging to Alessandro--Beatrice--Bertotti--Montorsi dynamics of propagating Portevin--Le Chatelier bands}

\author{Tero M\"akinen}
\affiliation{Department of Applied Physics, Aalto University, P.O. Box 15600, 00076 Aalto, Espoo, Finland}
\author{Mikko J. Alava}
\affiliation{Department of Applied Physics, Aalto University, P.O. Box 15600, 00076 Aalto, Espoo, Finland}

\date{\today}

\begin{abstract}
Type-A Portevin--Le Chatelier bands are propagating strain localizations whose velocity exhibits strong intermittent fluctuations. We derive a coarse-grained connection between the dynamic-strain-aging model of H\"ahner and Rizzi and the Alessandro--Beatrice--Bertotti--Montorsi (ABBM) mean-field model of avalanche dynamics. A propagating band is reduced to a single phase coordinate while retaining the elastic feedback of the tensile machine. This feedback generates the increasing drive and restoring terms of the ABBM description. A finite-time adjoint calculation determines the stress mobility of the band and is checked independently by direct stress continuation. 
The mobility is approximately proportional to the hardening parameter, whereas the band speed scales approximately as its square root. Short-range heterogeneity in the plastic strain deposited by the band is converted by the global machine feedback into a Brownian effective force. The resulting reduction provides a microscopic route from dynamic strain aging to ABBM-like avalanche dynamics of propagating PLC bands.
\end{abstract}

\maketitle

\section{Introduction}

The Portevin--Le Chatelier (PLC) effect~\cite{Lechatelier1909, Portevin1923} is a classical example of a macroscopic plastic instability in metallic solid solutions and alloys. Under suitable combinations of temperature and imposed deformation rate, otherwise smooth plastic flow becomes serrated and the strain localizes into inclined deformation bands~\cite{Yilmaz2011}. The standard microscopic picture is dynamic strain aging~(DSA): mobile dislocations are temporarily arrested at obstacles, solute atoms diffuse toward the arrested dislocations and increase their resistance to subsequent motion, and renewed glide partially destroys the aged configuration~\cite{CottrellBilby1949,McCormick1988}. The competition between the aging time and the characteristic time of plastic flow can generate negative asymptotic strain-rate sensitivity and an instability of homogeneous deformation.

The effect is conventionally classified into types A, B, and C. This is more than a classification of the serrations in the macroscopic stress signal: it also describes qualitatively distinct spatiotemporal modes of strain localization~\cite{Yilmaz2011,Jiang2007}. Type C bands nucleate at apparently unrelated positions, whereas type B deformation advances by spatially correlated but discontinuous hopping. Type A bands, the focus here, occur typically toward the high-strain-rate and low-temperature side of the window where the PLC effect is seen. A band nucleates in the gauge section and then propagates continuously over macroscopic distances, often through a substantial fraction of the specimen before it stops. This stop can be due to reaching the end of the gauge section, another band being nucleated, colliding with another band, or simply without any readily apparent reason. The global load signal contains relatively small and irregular stress drops, while optical measurements show a narrow region of intense plastic strain rate moving through an otherwise much less active gauge section~\cite{Chihab1987,Halim2007,AitAmokhtar2010}.
The band is sufficiently coherent to be represented to a good approximation by a time-dependent position~$X(t)$ and a velocity~$\dot X$.

Recent band-resolved experiments have substantially enlarged the quantitative data available for this picture. Digital image correlation~(DIC) has measured local PLC-band strains in austenitic stainless steel~\cite{Lee2021} in high temperature conditions; DIC under heterogeneous stress states has been used to track band nucleation and circumferential propagation in AA7075-W~\cite{Choi2021}; and high-speed DIC in AA5182 has measured band strain and velocity throughout individual tests and connected them to the local work-hardening state~\cite{XuBand2023,XuIJP2023}. Combined DIC and acoustic-emission measurements in AA5052 also emphasize that a serration type inferred from the macroscopic signal might not uniquely determine whether localization is continuously propagating or repeatedly nucleated~\cite{Liu2024}. 

This apparently continuous propagation does not imply microscopically smooth motion. High-speed speckle and image-correlation experiments have shown pulsations, short stopovers, and rapid local advances of type-A bands~\cite{Zhang2005,Xiang2007}. In AA5754~\cite{Makinen2020}, high-resolution speckle imaging showed that the instantaneous band velocity has large fluctuations organized into avalanches; their distributions and average avalanche shapes are well described by a mean-field crackling model, namely the 
Alessandro--Beatrice--Bertotti--Montorsi~(ABBM) model originally introduced for Barkhausen noise as a mean-field equation for a driven object moving through a Brownian pinning-force landscape~\cite{ABBM1,ABBM2,Cizeau1997}.
This combination---a stable mesoscale object with strongly fluctuating motion---makes type-A bands an unusually direct bridge between constitutive descriptions of DSA and statistical-physics descriptions of driven avalanches.

PLC-type serrations are not restricted to the traditional dilute Al--Mg and Cu--Al systems. Serration statistics resembling the conventional A--B--C regimes have been reported in low-, medium-, and high-entropy face-centered-cubic (FCC) alloy series~\cite{Carroll2015} and in Al$_{0.5}$CoCrFeNi~\cite{Niu2017}. More directly, DIC in the body-centered-cubic (BCC) HfNbTaTiZr system has tracked PLC-band propagation while the alloy is varied from low to high configurational entropy~\cite{Hsu2023}. Recent work on refractory HfNbTaTiZr connects serration statistics to slow-avalanche and fast-runaway dynamical regimes, both associated with the type-A part of the traditional PLC diagram over part of the phase space~\cite{Liu2025}. These systems broaden the materials context considerably: the microscopic solute-dislocation kinetics can be more complex than in a dilute binary alloy, but continuously propagating localization and avalanche-like dynamics remain experimentally relevant.

Two theoretical traditions are therefore naturally brought together. DSA-based constitutive models seek to describe the origin and spatiotemporal organization of the instability from dislocation–solute kinetics~\cite{Ananthakrishna2007,Hahner2002,HahnerRizzi2003,RizziHahner2004}. In contrast, the ABBM model was introduced as a mean-field avalanche model. In Ref.~\cite{Makinen2020}, the latter was adapted to a propagating PLC band and reproduced the measured avalanche statistics remarkably well. A later comparison with a modified Hähner–-Rizzi~(HR) model~\cite{Makinen2022} emphasized that a physically richer DSA description should, after coarse graining, approach an ABBM-like velocity dynamics if the two descriptions are to represent the same propagating state. Bridging the two traditions has been tried before: Hähner's two-part statistical treatment of DSA~\cite{hahner1996physics, hahner1996physics2} already cast the onset of unstable flow as a non-equilibrium phase transition. That mapping, however, concerns the threshold of the homogeneous instability rather than the motion of a band above it.

Here we make that connection explicit. We first summarize the ABBM and H\"ahner--Rizzi descriptions in Sec.~II. In Sec.~III we derive a collective-coordinate mapping for a type-A band and show that the testing-machine feedback produces the deterministic ABBM drive and restoring terms. Section~IV describes the numerical integration, phase extraction, finite-time adjoint, stress-continuation validation, strain-rate and finite-size sweeps, and quenched-disorder test. Section~V presents the resulting history, strain, rate, and size dependences and discusses what is, and is not, inherited by the coarse-grained ABBM parameters from the HR description. Section~VI summarizes the implications for conventional Al alloys and for newer alloy families in which type-A-like PLC dynamics is observed.

\section{Models}
\label{sec:models}

\subsection{ABBM model and its PLC-band formulation}

The ABBM model was introduced as a mean-field description of Barkhausen noise produced by a ferromagnetic domain wall moving through quenched disorder~\cite{ABBM1,ABBM2}. Its exact avalanche properties and its relation to Barkhausen experiments have been comprehensively studied~\cite{Colaiori2008, dobrinevski2012nonstationary, le2013avalanche}. In a convenient overdamped form, a collective spatial coordinate~$X$ (corresponding to the band position) is pulled by an increasing external drive (given by multiplying the time~$t$ with a driving constant~$c$) and opposed by a restoring force (Hookean, with a spring constant~$k$) and a random, spatially varying, pinning force~$F$. This gives the functional form
\begin{equation}
 \dot X=ct-kX+F(X). \label{eq:abbmX}
\end{equation}
The defining assumption is that the pinning force is Brownian in the coordinate~$X$,
\begin{equation}
 \left\langle [F(X)-F(X')]^2\right\rangle=2D|X-X'|, \label{eq:brownianDef}
\end{equation}
where $D$~fixes the disorder strength. This Brownian-force model arises naturally as the mean-field limit of an extended driven interface with many pinning contributions~\cite{Cizeau1997}. For monotonic motion, differentiating Eq.~(\ref{eq:abbmX}) with respect to time and using a time-change operation~\cite{baldassarri2021universal} to convert the stochastic term
(seeing that a Brownian process sampled along $X(t)$ acquires an increment proportional to $\sqrt{\dot X}$)
gives the form
\begin{equation}
 \dot v=c-kv+\sqrt{2Dv}\,\xi(t), \qquad v=\dot X, \label{eq:abbmV}
\end{equation}
with unit white noise $\xi(t)$. 
We note that this is also equivalent to the Cox--Ingersoll--Ross model~\cite{cox1985theory} used in finance.
The square-root multiplicative noise is responsible for the mean-field avalanche statistics.

For PLC bands, Ref.~\cite{Makinen2020} treated a type-A band as a rigid object whose position is obtained from imaging data. Equation~(\ref{eq:abbmX}) [or equivalently Eq.~(\ref{eq:abbmV})] was then used directly for the band coordinate. The parameters were not derived from a constitutive model: the driving coefficient was taken to be proportional to the applied strain rate $\edot$, $c=c_0\edot$ where $c_0$ is the proportionality constant, and the restoring coefficient proportional to the accumulated global strain $\epsilon$, $k=k_0\epsilon$ where $k_0$ is again the proportionality constant. The $D$ was kept constant, as the ABBM statistics depend only on the reduced quantities $c/D$ and $k/D$. With this minimal parametrization, the model reproduced the measured mean velocities and the statistics and shapes of velocity bursts. The experimentally determined mean band velocity in that work was approximately
\begin{equation}
 \langle v_{\rm b}\rangle\propto \edot^{p}\exp(-\epsilon/\epsilon_0),
 \qquad p\simeq0.6,
 \label{eq:expvb}
\end{equation}
where $\epsilon_0 = 0.16$ is the characteristic strain for the exponential decay. 
The present question is whether the structure and parameters of Eq.~(\ref{eq:abbmX}) can be generated from a DSA model rather than postulated.

\subsection{H\"ahner--Rizzi dynamic-strain-aging model}
\label{sec:HR}

The Hähner--Rizzi family of models, the original HR formulation~\cite{Hahner2002,HahnerRizzi2003,RizziHahner2004} and later closely related variants~\cite{Jiang2007}, belongs to a broader class of DSA descriptions in which the aging state is carried by a dynamical internal variable obeying its own relaxation equation, rather than being slaved instantaneously to the strain rate. The class originates with McCormick's effective aging time~\cite{estrin1991modelling, mccormick1995numerical}, and its members differ mainly in the choice of internal variable such as the aging time~\cite{McCormick1988} or the solute concentration on dislocation lines~\cite{Jiang2007}.
In the HR formulation this is the additional activation enthalpy.
The model includes both the stress-strain response, and the local behavior of the sample as a simple one-dimensional continuum representation. The starting point is thermally activated plastic flow where, following Ref.~\cite{Hahner2002}, the plastic strain $\ep$ evolves as
\begin{equation}
 \dot{\ep}=\nu \Omega
 \exp\left[-\frac{G_0+\Delta G}{k_{\rm B}T}
 +\frac{\sigma_{\rm eff}}{S_0}\right], \label{eq:arrhenius}
\end{equation}
where $\nu$ is an attempt frequency, $\Omega$ is the elementary plastic strain per activation cycle, $G_0$ is the activation enthalpy without DSA, $\Delta G$ is the additional aging enthalpy, the thermal energy scale is given by the Boltzmann constant $k_{\rm B}$ and the temperature~$T$, and $S_0$ is the instantaneous strain-rate sensitivity expressed as a stress scale. The effective stress is defined as the difference between the applied stress $\sigma_{\rm ext}$ and the internal stress $\sigma_{\rm int}$ (which includes the hardening effects), namely
\begin{equation}
 \sigma_{\rm eff}=\sigma_{\rm ext}-\sigma_{\rm int}. \label{eq:seff}
\end{equation}

The DSA variable $\Delta G$ increases while dislocations wait and is reduced by depinning. For the linear-aging case used in the numerical calculations below,
\begin{equation}
 \partial_t \Delta G=D_g\partial_x^2\Delta G
 +\eta(\Delta G_\infty-\Delta G)
 -\frac{\dot\ep}{\Omega}\Delta G, \label{eq:agingdim}
\end{equation}
where $\eta^{-1}$ is the aging time, $\Delta G_\infty$ the saturation value of the DSA variable, and $D_g$ the strength of a diffusion-like spatial coupling. In the original formulation this spatial term is a pseudodiffusion associated with correlated dislocation glide rather than ordinary solute diffusion. More general aging kinetics can be written with an exponent $n$; the linear case $n=1$ already contains the velocity-softening instability and was used for the compact analytical treatment of type-A propagation~\cite{Hahner2002}.
We note that the same aging-minus-depinning structure appears also in the McCormick aging-time~\cite{McCormick1988} and solute-concentration~\cite{Jiang2007} formulations, and in the coupled dislocation–solute model of Fressengeas \emph{et al.}~\cite{fressengeas2005dynamic}.

A key part of the model is the global elastic coupling to the tensile machine,
\begin{equation}
\dot{\sigma}_{\rm ext}
 =\Eeff\left(
 \edot_{\rm ext}-\overline{\dot\ep}
 \right), \label{eq:machine}
\end{equation}
where $\Eeff$ is the effective stiffness of specimen and testing machine, and
the overbar in $\overline{\dot\ep} = \frac{1}{L}\int_0^L\dot\ep(x,t)\,\mathrm{d}x$ denotes the spatial average over the gauge length~$L$. Plastic localization therefore unloads the entire gauge section. This global feedback will be central to the ABBM reduction.

For the linear hardening law of Ref.~\cite{Hahner2002}, we simplify matters by introducing the dimensionless notation
\begin{equation}
 g=\frac{\Delta G}{k_{\rm B}T},\qquad
 \tau=\eta t,
\end{equation}
with a natural extension to $g_\infty = \Delta G_\infty/(k_{\rm B}T)$,
and a generalized drive $f$ defined as
\begin{equation} \label{eq:fdef}
f = \frac{\nu}{\eta} \exp\left(
- \frac{G_0}{k_{\rm B} T} + \frac{\sigma_{\rm eff}}{S_0}
\right) .
\end{equation}
Keeping the dimensional coordinate~$x$, the equations can be written as
\begin{align}
 \partial_\tau f &= \frac{\mathrm{d}s}{\mathrm{d}\tau} f -\theta e^{-g}f^2, \label{eq:f}\\
 \partial_\tau g &= \frac{D_g}{\eta}\partial_x^2g+g_\infty-g-fe^{-g}g, \label{eq:g}
\end{align}
with
\begin{equation}
 \theta=\frac{\Omega H}{S_0}, \qquad
 \dot\ep=\eta \Omega f e^{-g}. \label{eq:theta_plastic}
\end{equation}
Here $H$ denotes the strain-hardening modulus and
$s=\sigma_{\rm ext}/S_0$ is the dimensionless applied stress,
with $\mathrm{d}s/\mathrm{d}\tau$ denoting the
dimensionless applied-stress rate.
The machine equation becomes
\begin{equation}
 \frac{\mathrm{d}s}{\mathrm{d}\tau} =\frac{\Eeff}{S_0}
 \left[\frac{v_{\rm cross}}{\eta L}-\Omega\overline{fe^{-g}}\right] .
 \label{eq:machineDimless}
\end{equation}

Equations~(\ref{eq:f}), (\ref{eq:g}), and (\ref{eq:machineDimless}) support the different PLC regimes and, in particular, continuously propagating type-A solutions. For a single band, the kinematics satisfy (by using the cross-head velocity $v_{\rm cross} = L \dot{\epsilon}_{\rm ext}$)
\begin{equation}
 v_{\rm cross}=\Deps\vb=\dot\epsilon_{\rm b}w_{\rm b}, \label{eq:solitaryKin}
\end{equation}
where $w_{\rm b}$, $\vb$, $\Deps$, and $\dot\epsilon_{\rm b}$ are the band width, propagation velocity, strain increment left by passage of the band, and the band strain rate. The HR papers derive type-A band parameters and show how the propagating solution loses stability toward other band modes~\cite{Hahner2002,HahnerRizzi2003,RizziHahner2004}.

A later version of the HR model was adapted to AA5754 data by using local Voce-like hardening, rate-dependent spatial coupling, and quenched variability in the initial (plastic strain rate) condition~\cite{Makinen2022}. This is, to our knowledge, the most direct recent use of the HR framework for quantitative type-A band dynamics. Finite-element modeling provides a complementary route in which DSA constitutive laws are embedded in the full specimen geometry. Early three-dimensional calculations based on McCormick-type kinetics reproduced propagating localized zones in Al--Mg--Si~\cite{ZhangFE2001}, and a study of AA5083 established the approach against matched experimental measurements~\cite{benallal2008experimental}.
More recent examples include a DSA-modified flow law for 20MnMoNi55 steel~\cite{Jana2019}, strain-aging viscoplastic simulations of highly reinforced AlCu/Al$_2$O$_3$ composites~\cite{Maziere2021}, McCormick-based three-dimensional simulations of the spatiotemporal PLC response in Ti--12Mo~\cite{Luo2023}, and large-strain thermoviscoplastic simulations of traveling instabilities in AW5083~\cite{Mucha2023}. The combined experimental--finite-element study of AA5182 by Xu~\emph{et~al.} is particularly relevant here because it resolves the band kinematics and uses a relation between gauge velocity, band strain, and band velocity~\cite{XuIJP2023}. Thus contemporary finite-element work generally uses McCormick-~\cite{McCormick1988}, Estrin--McCormick-~\cite{estrin1991modelling}, or related DSA laws rather than the HR equations themselves.
A recurring numerical finding in that literature is that DSA laws without an internal length give mesh-dependent band widths~\cite{maziere2010numerical}, motivating gradient or micromorphic regularization~\cite{maziere2015strain}. This role is played in HR formulation by the pseudodiffusion coefficient so HR remains an important analytical reference for band kinematics and propagation. The modified HR model reproduces the overall strain and strain-rate trends of type-A propagation, but the simple ABBM model remains superior for detailed avalanche statistics~\cite{Makinen2022}. In the present paper we deliberately use the original H\"ahner parameter set and hardening history. This keeps the mapping separate from a material-specific refit and allows comparison with the analytical scaling results of the original model.

\section{Collective-coordinate mapping}
\label{sec:mapping}

\subsection{Type-A band as a phase coordinate}

For a type-A band we define the local dimensionless activity
\begin{equation} \label{eq:activity}
 a(x,\tau)=f(x,\tau)e^{-g(x,\tau)}
\end{equation}
and the band phase is taken as the center of mass of this activity,
\begin{equation}
 X[\bm U]=\frac{\int x a(x)\,\mathrm{d}x}{\int a(x)\,\mathrm{d}x},
 \qquad \bm U=(f,g). \label{eq:phase}
\end{equation}
A rigid traveling state is then $\bm U(x,\tau)\simeq\bm U_0[x-X(\tau)]$ up to slow changes of band shape and background state. Translational invariance of an exact traveling solution gives the Goldstone mode $\bm U_0'$. H\"ahner \emph{et al.}~\cite{Hahner2002} explicitly identify the corresponding zero mode in their perturbation analysis of the type-A solution.

The passage of the band by $\mathrm{d}X$ leaves a plastic increment $\Deps$ over that part of the sample. To leading order,
\begin{equation}
 \mathrm{d}\overline{\ep}=\frac{\Deps}{L}\,\mathrm{d}X. \label{eq:dplastic}
\end{equation}
This is basically Eq.~(\ref{eq:solitaryKin}) written for the collective coordinate.

\subsection{Testing-machine feedback and the deterministic ABBM terms} \label{sec:feedback}

In the following, we use $\sigma = \sigma_{\rm ext}$ and $\epsilon = \epsilon_{\rm ext}$ for brevity.
Inserting Eq.~(\ref{eq:dplastic}) into Eq.~(\ref{eq:machine}) gives 
\begin{equation}
 \dot\sigma=\Eeff\left(\edot-\frac{\Deps}{L}\dot X\right), \label{eq:machineX}
\end{equation}
and, during a propagation interval where $\Deps$ changes slowly (stress starting from the initial value $\sigma_i$),
\begin{equation}
 \sigma(t)=\sigma_i+\Eeff\edot t-
 \frac{\Eeff\Deps}{L}X(t). \label{eq:sigmaX}
\end{equation}
The unloading responsible for the term linear in $X(t)$ has been observed directly. Speckle interferometry during a single band event in A2017 shows the elastic contraction of the rest of the gauge section, with the summed shrinkage measured as 6.46~µm compared with 6.223~µm from their accompanying simulation~\cite{Jiang2007}. Equation~(\ref{eq:sigmaX}) is the coarse-grained form of exactly this event-by-event global response.

Let the velocity of the propagating HR band respond to a small quasistatic stress change through its phase mobility~$\mus$, defined as
\begin{equation}
 \mus=\frac{\mathrm{d}\vb}{\mathrm{d}\sigma}. \label{eq:mu}
\end{equation}
Linearization about a reference propagation state (with stress~$\sigma^*$ and propagation velocity $\vb^*$)
gives $\dot X\simeq\vb^*+\mus(\sigma-\sigma^*)$. Substituting Eq.~(\ref{eq:sigmaX}) and absorbing constants into a shift of origins yields
\begin{equation}
 \dot X=ct-kX+\cdots, \label{eq:detMap}
\end{equation}
which looks like the ABBM equation
with
\begin{equation}
 c=\mus\Eeff\edot \label{eq:cmap}
\end{equation}
and
\begin{equation}
 k=\mus\Eeff\frac{\Deps}{L}. \label{eq:kmap}
\end{equation}
Thus the two deterministic terms of ABBM follow from the HR band mobility together with the global machine feedback. Neither $c$ nor $k$ is, in general, a bare microscopic parameter: both are renormalized by the state-dependent mobility and $k$ also contains the amount of strain deposited by the band.

For a single band satisfying Eq.~(\ref{eq:solitaryKin}), $\Deps=v_{\rm cross}/\vb$. Equations~(\ref{eq:cmap}) and (\ref{eq:kmap}) therefore imply the identity
\begin{equation}
 \frac{c}{k}=\vb. \label{eq:ckvb}
\end{equation}
This relation will be used as a numerical check.

\subsection{Adjoint phase response}
\label{sec:adjoint}

The mapping in Sec.~\ref{sec:feedback} requires one quantity that
cannot be read off the HR equations directly: the phase mobility
$\mu_\sigma$ [Eq.~(\ref{eq:musigma})], the sensitivity of the band
propagation speed to a small stress change. The obvious route is
direct continuation---perturb the stress, rerun the dynamics, and fit
the change in speed---and we use it as an independent check in
Sec.~\ref{sec:continuation}. Since $\mu_\sigma$ is needed at many
points of the loading history, however, it is preferable to obtain
the derivative from a single unperturbed trajectory. This is possible
because $\mu_\sigma$ is a linear response of a (nearly) translationally
invariant propagating state. 
For an ideal traveling wave, translational
invariance provides a zero mode of the linearized dynamics, the
Goldstone mode $\bm U_0'$ (explicitly identified for the type-A solution
in Ref.~\cite{Hahner2002}). Consider a generic scalar parameter~$\alpha$
entering the traveling-wave equations through the local evolution
operator $\bm R(\bm U;\alpha)$. Projecting the parameter-perturbed
traveling-wave equation onto the corresponding left zero mode
$\bm\Psi$ eliminates the unknown change of band shape and gives the
Fredholm solvability condition
\begin{equation}
\frac{\mathrm{d}V}{\mathrm{d}\alpha}
=-\frac{\langle\bm\Psi,\partial_\alpha\bm R\rangle}
        {\langle\bm\Psi,\bm U_0'\rangle}.
\label{eq:fredholm}
\end{equation}
Here $V$ is the dimensionless propagation speed and
$\langle\cdot,\cdot\rangle$ is the spatial inner product across the
band. The denominator fixes the normalization of the translational
mode, while the numerator measures the projection of the parameter
perturbation onto that mode. Equation~(\ref{eq:fredholm}) is the
standard traveling-wave phase-response relation; its derivation,
together with the definitions of the comoving linear operator and its
adjoint, is given in Appendix~\ref{app:adjoint}.

The stress response used in the numerical calculations is slightly
different from the generic parameter response in
Eq.~(\ref{eq:fredholm}). Consider a uniform perturbation of the
applied stress $\delta\sigma$ and define
$q=\delta\sigma/S_0$. In the fixed-stress calculation, $q$ does not
enter the subsequent evolution equations explicitly. Instead, because
$f$ depends exponentially on the applied stress
[Eq.~(\ref{eq:fdef})], it perturbs the initial state according to
\begin{equation}
f_q(x,0)=e^q f_0(x),\qquad
\left.
\frac{\partial\bm U_q(x,0)}{\partial q}
\right|_{q=0}
=
\begin{pmatrix}
f_0(x)\\
0
\end{pmatrix}.
\label{eq:qcoupling}
\end{equation}
Here $f_0(x)$ denotes the unperturbed field and $f_q(x,0)$ the field
immediately after applying the stress perturbation.

The Fredholm relation in Eq.~(\ref{eq:fredholm}) applies to a true
traveling wave with fixed asymptotic states and a parameter entering
the traveling-wave equations. The HR type-A band in the full tensile
test instead propagates on a slowly evolving, machine-driven
background, and the stress perturbation $q$ used here enters through
the initial condition in Eq.~(\ref{eq:qcoupling}). We therefore
evaluate the corresponding finite-time linear phase response with a
discrete adjoint of the actual fixed-stress trajectory. This
accumulates the sensitivity of the fitted phase speed to the
perturbed initial state in a single backward sweep. The implementation
is given in Sec.~\ref{sec:numerics} and validated against direct
continuation in Sec.~\ref{sec:validation}.

If $V$ is the fitted dimensionless phase velocity, the physical
mobility follows from the nondimensionalization. The HR spatial scale
is $\ell_0 = \sqrt{D_g/\eta}$, so a dimensionless phase speed
$V = \mathrm{d}(X/\ell_0)/\mathrm{d}\tau$ corresponds to
$v_\mathrm{b} = V\sqrt{D_g\eta}$, and therefore
$\mathrm{d}q = \mathrm{d}\sigma/S_0$. Hence
\begin{equation}
\mu_\sigma = \frac{\sqrt{D_g\eta}}{S_0}\,\frac{\mathrm{d}V}{\mathrm{d}q} .
\label{eq:musigma}
\end{equation}

\subsection{Short-range heterogeneity and the Brownian ABBM force}
\label{sec:disorderMap}

The deterministic HR equations do not, by themselves, specify a persistent quenched-disorder correlator. Consider instead weak spatial variability of the strain deposited by passage of the band,
\begin{equation}
 \Delta\epsilon_{\rm b}(X)=\overline{\Deps}+\delta \epsilon_{\rm b}(X), \label{eq:depsRandom}
\end{equation}
where $\delta \epsilon_{\rm b}$ is stationary and short-range correlated. The machine equation rewritten with this in mind integrates the plastic increment accumulated along the path,
\begin{equation}
 \sigma(t)=\sigma_i+\Eeff\edot t-
 \frac{\Eeff}{L}\int_0^{X(t)}
 \left[\overline{\Deps}+\delta \epsilon_{\rm b}(u)\right]\,\mathrm{d}u. \label{eq:sigmaRandom}
\end{equation}
The random contribution to the reduced velocity, i.e.~the missing term for the full ABBM equation, is therefore
\begin{equation}
 F_{\rm B}(X)=-\mus\frac{\Eeff}{L}\int_0^X\delta \epsilon_{\rm b}(u)\,\mathrm{d}u. \label{eq:FB}
\end{equation}
The quantity $F_{\rm B}$ is an effective force in the overdamped ABBM sense: after multiplication by the band mobility it has the dimensions of a velocity contribution, rather than of a mechanical force in newtons. Its physical origin is the fluctuating elastic unloading of the specimen. If a band passage deposits slightly more plastic strain than average in some spatial interval, the testing machine unloads more strongly and the subsequent band motion is retarded; if it deposits less, the elastic drive remains larger and the band is advanced relative to the mean trajectory. Equation~(\ref{eq:FB}) therefore represents the accumulated velocity bias generated by spatial variations of the plastic strain left in the wake of the band. It is this global accumulation, rather than the local heterogeneity alone, that produces a long-range random landscape for the coordinate $X$. 
We note that this is similar to the local hardening implemented in Ref.~\cite{Makinen2022} where the passage of a band leaves behind a changed material state.

If, for spatial separation $r$, $C_{\epsilon_{\rm b}}(r)=\langle\delta \epsilon_{\rm b}(x)\delta \epsilon_{\rm b}(x+r)\rangle$ has a finite integral, then for $|r|$ much larger than its correlation length,
\begin{equation}
 \left\langle[F_{\rm B}(X+r)-F_{\rm B}(X)]^2\right\rangle
 \simeq2D_{\rm eff}|r|, \label{eq:BrownianEmergence}
\end{equation}
with
\begin{equation}
 D_{\rm eff}=\frac12\left(\mus\frac{\Eeff}{L}\right)^2
 \int_{-\infty}^{\infty}C_{\epsilon_{\rm b}}(r)\,\mathrm{d}r. \label{eq:Dmap}
\end{equation}
Thus this Brownian force need not be microscopic. Ordinary finite-range heterogeneity becomes Brownian after spatial integration by the global elastic feedback. Direct local pinning contributions to the band coordinate may coexist with Eq.~(\ref{eq:FB}); the mapping identifies one explicit mechanism that produces the Brownian-force statistics assumed in the ABBM model on sufficiently long spatial scales.

\begin{table*}[tb!]
\caption{The parameters used in our simulations, which are identical to the ones used in the original Hähner \emph{et al.}~\cite{Hahner2002}.}
\label{tab:params}
\begin{ruledtabular}
\begin{tabular}{cccccccc}
$\eta$ & $\Omega$ & $S_0$ & $\Eeff$ & $D_g$ & $L$ & $v_{\rm cross}$ & $g_\infty$ \\
\hline
$0.1~{\rm s}^{-1}$ & $10^{-5}$ & $1~{\rm MPa}$ & $5 \times 10^4~{\rm MPa}$ & $10^{-7}~{\rm m^2/s}$ & $0.1~{\rm m}$ & $10^{-7}~{\rm m/s}$ & $6$\\
\end{tabular}
\end{ruledtabular}
\end{table*}

\section{Numerical approach}
\label{sec:numerics}

\subsection{Driven HR simulations}

We use the parameter set employed by H\"ahner \emph{et al.} for their type-A simulations \cite{Hahner2002}, shown in Table~\ref{tab:params}.
The natural spatial and velocity scales are
\begin{equation}
 \ell_0=\sqrt{D_g/\eta}=1~{\rm mm},
\end{equation}
and
\begin{equation}
 v_0=\sqrt{D_g\eta}=0.1~{\rm mm/s}.
\end{equation}
The gauge section is discretized into $N=100$ blocks of size $\ell_0$. The initial condition is $g=g_\infty$ and $f=10^{-13}$, with the second block multiplied by 30 to seed the first localization. Fixed end blocks are used, as in the original calculation. The hardening parameter $\theta$ is allowed to evolve through the instantaneous macroscopic hardening modulus $H$. H\"ahner \emph{et al.} did not keep $H$ constant in their type-A simulation; instead they prescribed a parabolic stress--strain response by allowing $\theta$ to decrease linearly with the spatially averaged accumulated plastic strain. We use that published prescription, by setting
\begin{equation}
 \theta(\overline\ep)=0.01-2\times10^{-6}\frac{\overline\ep}{\Omega}. \label{eq:thetaHist}
\end{equation}
Numerically it is advanced at every timestep
using $\mathrm{d}\overline\ep/\mathrm{d}\tau
=\Omega\,\overline{f e^{-g}}$, the spatial average of Eq.~(\ref{eq:theta_plastic}). With $\Omega=10^{-5}$ and $S_0=1$ MPa, the starting value $\theta_0=0.01$ corresponds to $H=1000$ MPa; over the type-A history analyzed below, the prescribed hardening modulus decreases substantially, as in Ref.~\cite{Hahner2002}.

The full driven equations are integrated by explicit Euler method with $\Delta\tau=0.025$ to $\tau=4500$ and sampled every $0.5$ in $\tau$. Candidate type-A snapshots are retained only if the activity maximum lies between blocks 30 and 70, a local linear fit to the position of the activity maximum over a $\pm5$ interval has a root-mean-square residual below 0.4 blocks, and the absolute dimensionless speed lies between 0.2 and 1.5. Up to five approximately evenly spaced clean snapshots are used in each plastic-strain bin of width 0.005.

The phase in Eq.~(\ref{eq:phase}) and the position of the activity maximum give consistent speeds for these clean traverses.
The deposited strain is obtained independently from the single-band kinematic relation of Eq.~(\ref{eq:solitaryKin}).

\subsection{Finite-time discrete adjoint}

At each selected snapshot the external stress is frozen ($\mathrm{d}s/\mathrm{d}\tau=0$), and Eqs.~(\ref{eq:f}) and (\ref{eq:g}) are evolved over $0\le\tau\le16$ with $\Delta\tau=0.01$. 
The dimensionless speed $V$ is the least-squares slope of $X(\tau)/\ell_0$ over $2\le\tau\le10$.
Discarding $\tau < 2$ is not merely a transient cut for the phase fit. A stress step acts first through the instantaneous sensitivity $S_0$---Eq.~(\ref{eq:qcoupling}) is precisely the statement that $f$ responds instantaneously and exponentially to $\delta \sigma$ while $g$ does not---and the aging field only redistributes over a retardation time of order the aging time $\eta^{-1}$~\cite{hahner1996physics}. With $\eta = 0.1$~s$^{-1}$ the fitting window corresponds to two to ten aging times after the perturbation, so $\mu_{\sigma}$ as extracted here is an asymptotic rather than an instantaneous mobility.

Let $\bm{\mathcal{G}}(\bm U;\theta)$ denote the complete fixed-stress
right-hand side of Eqs.~(\ref{eq:f}) and (\ref{eq:g}), including the
spatial-coupling term in the $g$ component. One explicit-Euler step
can then be written as
\begin{equation}
\bm U_{n+1}
=\bm U_n+\Delta\tau\,\bm{\mathcal{G}}(\bm U_n;\theta)
\equiv\bm\Phi(\bm U_n).
\label{eq:euler}
\end{equation}
The fitted slope is the linear functional $V=\sum_nw_nX(\bm U_n)$, where $w_n$ are the fixed least-squares slope weights. 
Introducing the discrete adjoint variable $\bm\lambda_n$, the backward recursion obeys
\begin{equation}
 \bm\lambda_n=\left(\frac{\partial\bm\Phi}{\partial\bm U_n}\right)^T
 \bm\lambda_{n+1}+w_n\nabla_{\bm U}X(\bm U_n). \label{eq:adjoint}
\end{equation}
Using Eq.~(\ref{eq:qcoupling}),
\begin{equation}
 \frac{\mathrm{d}V}{\mathrm{d}q}=\bm\lambda_0\cdot
 \begin{pmatrix}f_0\\0\end{pmatrix}. \label{eq:dCdq}
\end{equation}
One fixed-stress forward trajectory and one backward sweep therefore give the stress mobility without a family of perturbed simulations.

For the discrete phase, with $A=\sum_i a_i$ and $X=\sum_ix_ia_i/A$,
\begin{equation}
 \frac{\partial X}{\partial f_i} =\frac{x_i-X}{A}e^{-g_i},
\end{equation}
and
\begin{equation}
 \frac{\partial X}{\partial g_i} =-\frac{x_i-X}{A}f_ie^{-g_i}.
\end{equation}
The local reaction contribution to the Jacobian used in the adjoint is
\begin{equation}
 J_{\rm reac}=\begin{pmatrix}
 -2\theta f e^{-g} & \theta f^2e^{-g}\\
 -ge^{-g} & -1-fe^{-g}(1-g)
 \end{pmatrix}, \label{eq:Jac}
\end{equation}
combined with the transpose of the second-difference operator in the $g$ component.

We also check the effect of the timestep size, by setting $\Delta\tau=0.02$, 0.01, and 0.005, and get for an early clean band $(V,\mathrm{d}V/\mathrm{d}q)=(0.93569,1.87199)$, $(0.93733,1.87957)$, and $(0.93816,1.88454)$, respectively. Thus halving the timestep changes the phase speed by less than 0.1\% and the phase response by about 0.3\% in this representative state.

\subsection{Direct continuation and strain-rate sweeps} \label{sec:continuation}

The adjoint derivative is checked by applying $q=-0.005,-0.0025,0,0.0025,0.005$ to the same initial snapshot, evolving the fixed-stress dynamics, and fitting $V(q)$. We use states from early, intermediate, and late parts of the type-A history.

To probe the imposed-rate dependence while keeping the hardening state nearly fixed, we also run first traverses of the band at $\edot=5.0\times10^{-7}$, $7.5\times10^{-7}$, $1.0\times10^{-6}$, and $1.25\times10^{-6}$~s$^{-1}$. The first clean central part of the traverse is analyzed, for which the accumulated plastic strain is below $4.1\times10^{-4}$ and $\theta$ differs by less than 0.2\% among the four simulations.

To probe finite-size effects, we repeat the early-history calculation for gauge lengths $L=50$, 75, 100, 125, and 150~mm while keeping the local block size fixed at 1~mm, the nominal strain rate fixed at $\edot=10^{-6}$~s$^{-1}$, and all constitutive parameters unchanged. Thus the number of blocks changes with $L$ and the crosshead velocity is changed consistently as $v_{\rm cross}=L\edot$. The first clean central traverse is analyzed in each case using the same activity-peak criterion as above. A 200-mm simulation was also performed, but it did not satisfy the unchanged clean single-band selection criterion over the same analysis window and is therefore not included in the fitted finite-size exponents.

\subsection{Quenched-disorder test}

To test Eq.~(\ref{eq:Dmap}) explicitly, we introduce a weak persistent local activation-barrier field $\zeta_i$ through
\begin{equation}
 a_i=f_i\exp[-g_i-\zeta_i], \label{eq:zeta}
\end{equation}
with independent zero-mean Gaussian $\zeta_i$ of standard
deviation $\sigma_\zeta$ on each 1-mm block.
This is an explicit extension used for the coarse-graining test, not a term in the original HR equations. We use $\sigma_\zeta=0.01$, 0.02, 0.03, and 0.04 and 20 independent realizations per amplitude. The deposited strain after the first passage is accumulated on blocks 15--85. After subtracting the ensemble-mean spatial profile, the cumulative force in Eq.~(\ref{eq:FB}) is formed using the early-history adjoint mobility $\mus=0.2097$~mm/(s MPa). The mean-square force increment is fitted for separations 10--25~mm, well above the one-block disorder correlation length.

\begin{figure}[t]
\centering
\includegraphics[width=\columnwidth]{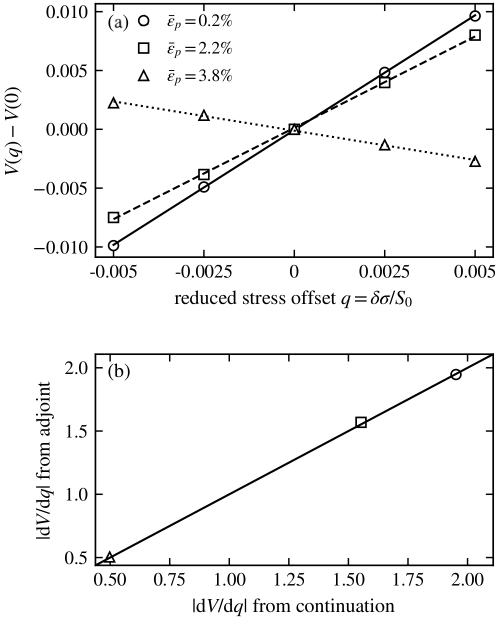}
\caption{Validation of the finite-time discrete adjoint. (a)~Direct continuation of the fitted dimensionless band speed $V$ under small stress offsets $q$ for three representative states. Lines are linear fits. (b)~Adjoint derivative versus the derivative from direct continuation; the solid line denotes equality.}
\label{fig:adjointvalidation}
\end{figure}

\section{Results and exploration}
\label{sec:results}

\begin{table}[tb]
\caption{Validation of the finite-time adjoint by direct continuation in
$q$, where ``adj'' and ``cont'' denote the adjoint and
direct-continuation estimates, respectively.}
\label{tab:cont}
\begin{ruledtabular}
\begin{tabular}{cccc}
$\overline\ep$ & $(\mathrm{d}V/\mathrm{d}q)_{\rm adj}$ & $(\mathrm{d}V/\mathrm{d}q)_{\rm cont}$ & $R^2$\\
\hline
0.00234 & 1.9469 & 1.9514 & 0.99994\\
0.02179 & 1.5688 & 1.5529 & 0.99961\\
0.03791 & $-0.5046$ & $-0.4981$ & 0.99692\\
\end{tabular}
\end{ruledtabular}
\end{table}

\subsection{Adjoint validation} \label{sec:validation}

The purpose of this validation is numerical: the adjoint computes the derivative of the fitted band speed with respect to an infinitesimal stress perturbation from a single unperturbed trajectory, whereas direct continuation obtains the same derivative by rerunning the fixed-stress dynamics at several small positive and negative stress offsets and fitting the resulting $V(q)$ curve. The two calculations use the same phase definition but otherwise evaluate the response independently. Agreement therefore checks the discrete-adjoint implementation, the phase gradient, and the assumed linear-response window. Table~\ref{tab:cont} gives this comparison. The $V(q)$ response is almost perfectly linear across the chosen perturbation interval and the derivative agrees within about 1.3\% for all three states. The sign in the late state reflects the direction of propagation; history averages below use a positive mobility convention for forward motion.

Figure~\ref{fig:adjointvalidation} visualizes the same check. First, Fig.~\ref{fig:adjointvalidation}a shows the continuation curves $V(q)$ for representative early-, intermediate-, and late-history states. Then, Fig.~\ref{fig:adjointvalidation}b compares the slope extracted from continuation with the derivative obtained from the adjoint, and shows a near-diagonal collapse. Here ``validation'' means validation of the numerical phase-response calculation, not an independent validation of the constitutive model itself.

\subsection{History dependence and hardening}

\begin{figure}[t]
\centering
\includegraphics[width=\columnwidth]{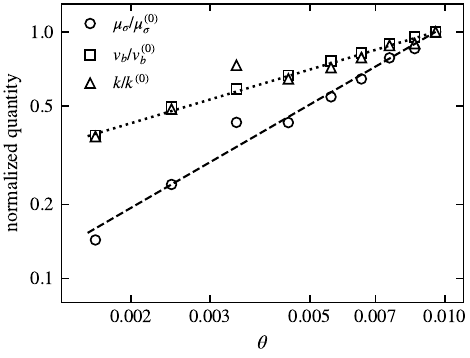}
\caption{History dependence of the phase mobility, band velocity, and coarse-grained stiffness, normalized by their first-bin values. Open symbols denote the numerical data. The black dashed and dotted lines show the fits $\theta^{1.052}$ and $\theta^{0.546}$, respectively.}
\label{fig:history}
\end{figure}

Table~\ref{tab:history} gives the phase mobility and the corresponding ABBM coefficients extracted through the simulated loading history. The mobility decreases strongly as the published hardening coefficient $\theta$ decreases. 
Logarithmic fits over the nine bins give the following scaling
relations, where $R^2_{\log}$ denotes the coefficient of
determination evaluated in logarithmic coordinates:
\begin{align}
 \mus\propto\theta^{1.052},\qquad &\textrm{with} \: R^2_{\log}=0.977, \label{eq:mufit}\\
 \vb\propto\theta^{0.546},\qquad &\textrm{with} \: R^2_{\log}=0.998, \: \textrm{and} \label{eq:vbfit}\\
 k\propto\theta^{0.506},\qquad &\textrm{with} \: R^2_{\log}=0.920. \label{eq:kfit}
\end{align}
The velocity exponent is essentially the $0.54$ hardening exponent reported independently in the original H\"ahner simulations. The new result is the approximately linear dependence of the stress phase mobility on $\theta$. Recent DIC measurements on AA5182 provide a useful experimental counterpart: Xu \emph{et al.} extracted the band strain and velocity event by event and found $v_{\rm b}/v_{\rm cross}\propto\Theta$, where $\Theta$ is the instantaneous material work-hardening rate \cite{XuBand2023,XuIJP2023}. Their $\Theta$ is not identical to the dimensionless HR coefficient $\theta$, so the exponents should not be equated directly; nevertheless, both results identify the evolving hardening state as a central control on type-A kinematics. Since at fixed imposed rate $\Deps\propto1/\vb$, Eq.~(\ref{eq:kmap}) predicts
\begin{equation}
 k\propto\frac{\mus}{\vb}\sim\theta^{1.052-0.546}
 =\theta^{0.506},
\end{equation}
in agreement with the directly extracted $k$ exponent.
Figure~\ref{fig:history} shows these history-dependent trends and makes the different powers of the same evolving hardening parameter directly comparable.

\begin{table*}[t]
\caption{Collective-coordinate reduction through the H\"ahner loading history. $\mus$ is the median of up to five clean-band snapshots in each strain bin; the uncertainty is the snapshot-to-snapshot standard deviation.}
\label{tab:history}
\begin{ruledtabular}
\begin{tabular}{ccccccc}
$\overline\ep$ & $\theta$ & $\vb$ (mm/s) & $\mus$ (mm/(s MPa)) & $\Deps$ & $c$ (mm/s$^2$) & $k$ (s$^{-1}$)\\
\hline
0.00235 & 0.009530 & 0.10234 & $0.2097\pm0.0166$ & 0.000977 & 0.01049 & 0.1025\\
0.00720 & 0.008559 & 0.09766 & $0.1794\pm0.0123$ & 0.001024 & 0.00897 & 0.0918\\
0.01239 & 0.007522 & 0.09117 & $0.1645\pm0.0134$ & 0.001097 & 0.00823 & 0.0902\\
0.01743 & 0.006513 & 0.08390 & $0.1351\pm0.0104$ & 0.001192 & 0.00676 & 0.0805\\
0.02213 & 0.005574 & 0.07792 & $0.1143\pm0.0272$ & 0.001283 & 0.00572 & 0.0734\\
0.02761 & 0.004478 & 0.06805 & $0.0899\pm0.0253$ & 0.001469 & 0.00449 & 0.0660\\
0.03284 & 0.003432 & 0.06000 & $0.0901\pm0.0273$ & 0.001667 & 0.00450 & 0.0751\\
0.03770 & 0.002460 & 0.05065 & $0.0505\pm0.0134$ & 0.001974 & 0.00252 & 0.0498\\
0.04167 & 0.001665 & 0.03896 & $0.0301\pm0.0325$ & 0.002567 & 0.00150 & 0.0386\\
\end{tabular}
\end{ruledtabular}
\end{table*}

\begin{figure}[t]
\centering
\includegraphics[width=\columnwidth]{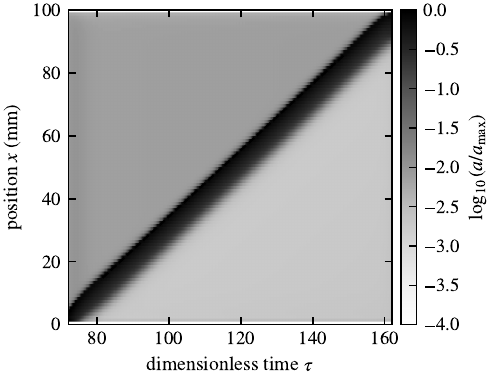}
\caption{Space--time map of the normalized plastic activity $a$ [Eq.~(\ref{eq:activity})] during one simulated type-A traverse of the H\"ahner--Rizzi model. A linear fit to the activity maximum over the central part of the traverse corresponds to a physical speed of approximately $0.106$~mm/s.}
\label{fig:propagation}
\end{figure}

A representative clean traverse from the same driven simulation is shown in Fig.~\ref{fig:propagation}. The activity remains localized in a band of finite width while its center advances almost linearly through the gauge section. 
This confirms the expected behavior of a band in a medium without added disorder.
A fit to the activity maximum over the central part of this passage gives $\vb\simeq0.106$~mm/s, consistent with the speed used in the collective-coordinate extraction. The figure also illustrates why a single phase coordinate is meaningful on propagation timescales even though the local activity inside the band is spatially extended.

The consequence is important for the phenomenological PLC--ABBM parametrization. At fixed $\edot$,
$
 c(\epsilon)=\Eeff\edot\,\mus(\epsilon),
$ 
so the driving coefficient is not state independent in the HR reduction. Similarly, $k$ is not simply the microscopic work-hardening modulus and does not have to increase linearly with accumulated strain. The choices $c\propto\edot$ and $k\propto\epsilon$ in Ref.~\cite{Makinen2020} should therefore be interpreted as compact effective parametrizations over the experimental window rather than direct identifications with HR variables.

\subsection{Dependence on accumulated strain}

\begin{figure}[t]
\centering
\includegraphics[width=\columnwidth]{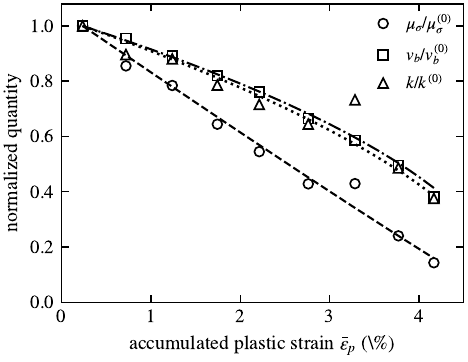}
\caption{Dependence on accumulated plastic strain for the HR loading history. Symbols are the numerical phase mobility, type-A band velocity, and coarse-grained stiffness normalized by their first-bin values. Curves are obtained by combining Eq.~(\ref{eq:thetaeps}) with the fitted hardening exponents.}
\label{fig:strain}
\end{figure}

The same loading history can be expressed directly in terms of the accumulated plastic strain, a quantity that is more directly accessible experimentally than the internal HR hardening parameter. With $\Omega=10^{-5}$, Eq.~(\ref{eq:thetaHist}) becomes
\begin{equation}
 \theta(\overline\ep)=0.01-0.2\,\overline\ep. \label{eq:thetaeps}
\end{equation}
Combining Eq.~(\ref{eq:thetaeps}) with Eqs.~(\ref{eq:mufit})--(\ref{eq:kfit}) gives explicit strain-dependent predictions for the coarse-grained coefficients. Figure~\ref{fig:strain} shows the numerical values against accumulated plastic strain together with the curves obtained by inserting Eq.~(\ref{eq:thetaeps}) into the fitted hardening power laws. Across the simulated interval, 0.2--4.2\% plastic strain, the normalized phase mobility decreases to about 0.14 of its early-history value, while the band velocity and effective stiffness both decrease to roughly 0.38. These expressions are intended only for the simulated strain interval and are not extrapolated toward the point where the prescribed $\theta$ would vanish.

\subsection{Early-history strain-rate dependence}

The rate sweep is summarized in Table~\ref{tab:rate}. Logarithmic fits give
\begin{equation}
 \vb\sim\edot^{0.697},\qquad
 \mus\sim\edot^{0.670}, \label{eq:rateFits}
\end{equation}
and consequently $c\sim\edot^{1.670}$ and $k\sim\edot^{0.973}$ over this restricted interval. The HR band-speed exponent is of the same order as the experimental $p\simeq0.6$ in Eq.~(\ref{eq:expvb}), but the mobility fit has only $R^2_{\log}=0.877$. We therefore regard these rate exponents as exploratory rather than asymptotic scaling laws. The stronger statement is that the constitutive band mobility itself changes with driving rate, so a reduction that sets $c$ exactly proportional to $\edot$ is not generally expected outside a narrow regime.

\begin{table}[tb]
\caption{Early-history strain-rate sweep of the H\"ahner--Rizzi model at nearly fixed hardening state.}
\label{tab:rate}
\begin{ruledtabular}
\begin{tabular}{ccccc}
$\edot$ (s$^{-1}$) & $\vb$ & $\mus$ & $c$ & $k$\\
 & (mm/s) & (mm/(s MPa)) & (mm/s$^2$) & (s$^{-1}$)\\
\hline
$5.0\times10^{-7}$ & 0.0646 & 0.1364 & 0.00341 & 0.0528\\
$7.5\times10^{-7}$ & 0.0868 & 0.1554 & 0.00583 & 0.0672\\
$1.0\times10^{-6}$ & 0.1052 & 0.1853 & 0.00926 & 0.0881\\
$1.25\times10^{-6}$ & 0.1225 & 0.2621 & 0.01638 & 0.1337\\
\end{tabular}
\end{ruledtabular}
\end{table}

\begin{figure*}[t]
\centering
\includegraphics[width=\textwidth]{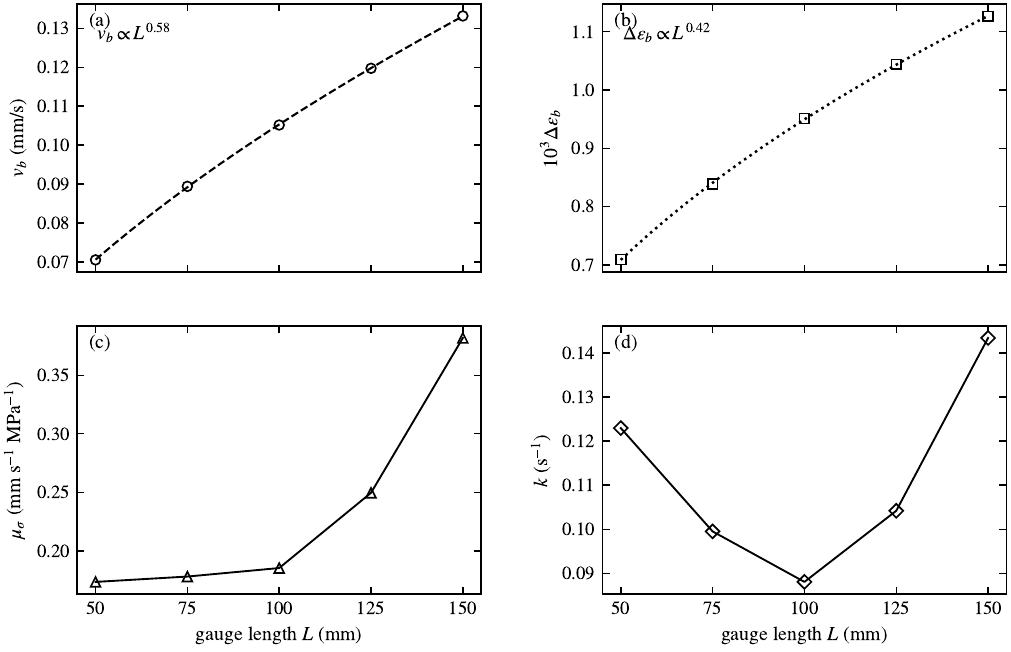}
\caption{Finite-size dependence at fixed nominal strain rate $\edot=10^{-6}$~s$^{-1}$ and fixed 1-mm spatial resolution. 
(a)~Band velocity versus gauge length. 
(b)~Plastic strain increment deposited by the band versus gauge length. Dashed and dotted lines in panels (a) and (b) are power-law fits over $L=50$--150~mm. 
(c)~Phase mobility $\mu_\sigma$ versus gauge length.
(d)~Effective ABBM restoring coefficient $k$ versus gauge length. 
Panels (c) and (d) do not display a simple finite-size scaling over the present range.}
\label{fig:size}
\end{figure*}

\subsection{Finite-size effects}

The gauge-length sweep gives a clean finite-size dependence of the band kinematics while the constitutive parameters, nominal strain rate, and 1-mm local resolution are held fixed. 
Figure~\ref{fig:size} summarizes the four collective quantities extracted from the reduction. Power-law fits over $L=50$--150~mm give
\begin{align}
 \vb \propto L^{0.577}, \qquad &\textrm{with} \: R^2_{\log}=0.99996, \: \textrm{and}\label{eq:sizevb}\\
 \Deps \propto L^{0.423}, \qquad &\textrm{with} \: R^2_{\log}=0.99993 \label{eq:sizedeps}
\end{align}
shown in Fig.~\ref{fig:size}a-b.
The exponents sum to unity within numerical precision, as required by the single-band kinematic identity $L\edot=\Deps\vb$ at fixed nominal strain rate. The size dependence is therefore shared between the propagation speed and the local strain deposited by the passing band rather than appearing solely as a proportional increase of $\vb$ with gauge length.

By contrast, the finite-size dependence of the phase mobility $\mus$ and of the resulting ABBM restoring coefficient $k$ do not look like power-laws (see Fig.\ref{fig:size}c-d) in the present range of system sizes. Thus the naive expectation $k \propto L^{-1}$, which would follow if both the local mobility and deposited strain were independent of system size, is not supported by these simulations. This is useful for the coarse-graining interpretation because it shows that the effective ABBM coefficients can inherit global geometric information in addition to the local constitutive state.

\begin{table}[tb]
\caption{The effective strengths $D_{\rm eff}$ generated by the local activation-barrier disorder [Eq.~(\ref{eq:zeta})] for different disorder amplitudes~$\sigma_\zeta$.}
\label{tab:disorder}
\begin{ruledtabular}
\begin{tabular}{c|cccc}
$\sigma_\zeta$ &0.01&0.02&0.03&0.04\\ \hline
$D_{\rm eff}$ &$2.26\!\times\!10^{-7}$&$7.60\!\times\!10^{-7}$&$2.07\!\times\!10^{-6}$&$4.01\!\times\!10^{-6}$\\
\end{tabular}
\end{ruledtabular}
\end{table}

\subsection{Emergent Brownian disorder}

The explicit disorder test produces the effective strengths shown in Table~\ref{tab:disorder}. The mean-square increments grow linearly with spatial separation on the fitting interval. Across the disorder amplitudes,
$
 D_{\rm eff}\propto\sigma_\zeta^{2.08},
$ 
and a constrained quadratic fit gives
\begin{equation}
 D_{\rm eff}\simeq2.43\times10^{-3}\sigma_\zeta^2. \label{eq:Dnum}
\end{equation}
The absolute coefficient depends on the force and spatial normalization and should not be compared directly with the $D$ fitted in Ref.~\cite{Makinen2020}. The relevant result is the Brownian-in-distance growth and the expected quadratic dependence on weak microscopic disorder.

\begin{figure}[t]
\centering
\includegraphics[width=\columnwidth]{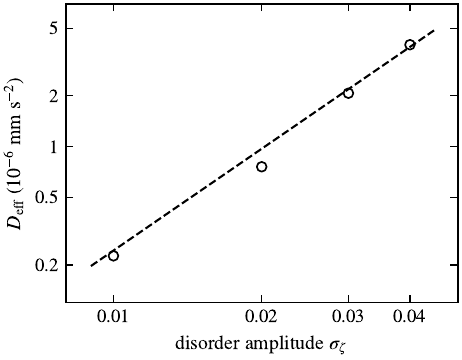}
\caption{Effective Brownian-force strength generated by weak persistent barrier disorder. Open symbols are numerical results and the black dashed line is the constrained quadratic fit in Eq.~(\ref{eq:Dnum}).}
\label{fig:disorder}
\end{figure}

\subsection{What is inherited from HR and what is effective?}

The calculation supports the hierarchy
\begin{equation}
 (f,g,\sigma)\ \longrightarrow\ X(t)
 \ \longrightarrow\ \dot X=ct-kX+F(X) \label{eq:hierarchy}
\end{equation}
which means going from the HR parameters to the band position, and deriving the ABBM equation for the band position.
The deterministic ABBM structure is not imposed after the fact: the ramp $ct$ and restoring term $-kX$ are consequences of elastic loading and the integrated plastic relaxation. The Brownian random force also has a concrete coarse-graining route through short-range spatial variability of the deposited strain. What is not universal are the coefficients. The mobility $\mus$, the deposited strain $\Deps$, and the microscopic disorder correlator all depend on the constitutive state and on material parameters. Thus $c$, $k$, and $D$ should in general be history-, rate-, temperature-, and material-dependent effective quantities.

The reduction also clarifies why the simple ABBM model can capture avalanche statistics even when a more detailed DSA model is needed if one is interested in the actual stress-strain behavior or wants to predict where in parameter space type-A propagation exists. The HR model contains the aging time, the instability window, hardening, spatial coupling, and transitions among band types. Once a coherent type-A band exists, however, many of these internal variables affect its long-scale motion primarily through the mobility, the deposited strain, and the statistics of the force landscape encountered by the translational mode. ABBM discards the internal band structure but retains precisely these ingredients at the collective-coordinate level.

\section{Conclusions}
\label{sec:conclusions}

We have derived and numerically tested a collective-coordinate mapping from the H\"ahner--Rizzi DSA description of a propagating type-A PLC band to ABBM dynamics.
The global tensile-machine equation converts the motion of a band
into a linearly increasing drive and
an elastic restoring force. The resulting coefficients are given by
Eqs.~(\ref{eq:cmap}) and (\ref{eq:kmap}) for a single propagating band.
A finite-time discrete adjoint provides the required stress mobility and agrees with direct stress continuation.

For the original H\"ahner type-A parameters, the loading history gives $\mus\sim\theta^{1.052}$ and $\vb\sim\theta^{0.546}$, which in turn produces $k\sim\theta^{0.506}$. The result makes clear that the effective ABBM parameters need not follow the simple strain and rate dependences used in the earlier phenomenological fit to AA5754. Those choices successfully summarize an experimental window, while a constitutive reduction predicts that the coefficients are renormalized by the instantaneous state of the DSA band. A weak quenched barrier field further demonstrates that short-range fluctuations of deposited strain are transformed by the machine feedback into a Brownian force, providing one explicit microscopic origin of the ABBM noise.

The additional parameter sweeps show that this renormalization is not only a function of the internal hardening state. Expressing the history directly in terms of accumulated strain reveals substantial evolution of the effective mobility, velocity, and restoring coefficient over only a few percent plastic strain. Varying the gauge length at fixed nominal strain rate gives $\vb\sim L^{0.577}$ and $\Deps\sim L^{0.423}$, while $\mus$ and $k$ do not follow simple finite-size power laws. The effective ABBM coefficients should therefore be viewed as coarse-grained quantities that inherit both constitutive history and specimen/loading geometry.

The mapping suggests several direct experimental tests. Simultaneous high-speed imaging and small controlled load perturbations could determine a phase mobility $\mathrm{d}\vb/\mathrm{d}\sigma$ independently of avalanche fitting. The strain increment deposited behind an individual type-A band can be measured by DIC and inserted into Eq.~(\ref{eq:kmap}). Their product then predicts the restoring coefficient governing velocity relaxation. These measurements would separate material-specific DSA kinetics from the more general collective-coordinate dynamics.

The avalanche character of type-A propagation is also mechanically relevant rather than merely a statistical refinement of a smooth band picture. The macroscopic band advances continuously over long distances, but the plastic strain rate and the associated stress relaxation are concentrated into bursts on shorter time and length scales~\cite{Zhang2005,Xiang2007,Makinen2020}. Their size and duration distributions~\cite{lebyodkin2000spatio} therefore control how a nominally steady band partitions plastic strain in space and time. In a PLC material this intermittency contributes directly to stress serrations, local strain accumulation, and the fluctuations of the effective work-hardening response; through repeated passages it can also influence where persistent localization develops before necking or failure. In the present mapping, the ABBM disorder parameter and the state-dependent mobility provide quantitative measures of these fluctuations. Establishing how avalanche statistics change with alloy chemistry, hardening state, and machine stiffness is consequently relevant to the mechanical response and formability of materials that exhibit PLC, not only to the universality class of their crackling noise.

Such tests need not be limited to simple alloys like Al--Mg. Recent DIC-based work has resolved propagating or spatially organized PLC localization in austenitic stainless steel~\cite{Lee2021}, Ni--Co superalloys~\cite{LiuNiCo2020}, compositionally complex HfNbTaTiZr alloys~\cite{Hsu2023}, and high-Mn austenitic steel~\cite{Hwang2026}. The last of these directly measures the local strain-hardening response associated with propagating PLC bands, which is especially relevant to the present identification of hardening-controlled mobility. In Inconel 718, a combined DIC--finite-element study explicitly includes testing-machine stiffness in the analysis of spatiotemporal PLC dynamics~\cite{Billon2026}, highlighting the same global feedback that generates the $ct-kX$ structure here. Avalanche-based phase diagrams of refractory high-entropy alloys provide a complementary statistical view~\cite{Liu2025}. The microscopic aging and weakening mechanisms in these materials may differ in detail from dilute binary alloys, and the applicability of the HR constitutive form itself should not be assumed. The coarse-grained prediction is weaker and more testable: whenever deformation organizes into a coherent propagating band whose motion unloads a globally coupled specimen, an ABBM-like coordinate equation is a natural candidate at long scales. How the effective $c$, $k$, and $D$ vary across alloy chemistry, temperature, machine stiffness, and band-type boundaries then becomes a quantitative way to compare distinct microscopic PLC mechanisms.

\begin{acknowledgments}
M.J.A. acknowledges support from the Academy of Finland Center of Excellence program (Grants No.~278367 and 317464) and the Finnish Cultural Foundation. 
T.M. acknowledges support from the Research Council of Finland (Grant No.~130265).
T.M. and M.J.A. acknowledge support from the FinnCERES flagship (Grant No.~151830423), Business Finland (Grants No.~211835, 211909, and 211989), the Research Council of Finland (Grant No.~13361245), and Future Makers programs. The authors acknowledge the computational resources provided by the Aalto University School of Science ``Science-IT'' project.
\end{acknowledgments}

\appendix

\section{Derivation of the phase-response formula}
\label{app:adjoint}

Here we derive the generic Fredholm phase-response relation
Eq.~(\ref{eq:fredholm}) and define the operators appearing in it.
The scalar parameter $\alpha$ introduced in Sec.~\ref{sec:adjoint}
enters the local evolution equations of an ideal traveling wave, and we write these equations
compactly as
\begin{equation}
\partial_\tau\bm U
=\mathcal D\,\partial_x^2\bm U+\bm R(\bm U;\alpha),
\label{eq:compact}
\end{equation}
where $\mathcal D$ is the diffusion matrix, acting only on the
$g$ component with coefficient $D_g/\eta$, and $\bm R$ denotes the
local reaction terms.
For a state propagating rigidly at
dimensionless speed $V$, substituting $\bm U(x,\tau)=\bm U_0(z)$ with
$z = x - V\tau$ into Eq.~(\ref{eq:compact}) gives the profile
equation
\begin{equation}
\mathcal{D}\,\bm U_0'' + V\,\bm U_0'
+\bm R(\bm U_0;\alpha)=0
\label{eq:twprofile}
\end{equation}
where primes denote derivatives with respect to the comoving coordinate $z$.\\

To do the linearization and to define the zero mode, we 
let $(\bm U_0^*,V^*)$ solve Eq.~(\ref{eq:twprofile}) at a reference
parameter value $\alpha_0$. Expanding
$\bm U_0\to\bm U_0^*+\delta\bm U$,
$V\to V^*+\delta V$, and
$\alpha\to \alpha_0+\delta \alpha$ to first order gives
\begin{equation}
\mathcal L\,\delta\bm U
+\bm U_0^{*\prime}\,\delta V
+\left.
\frac{\partial\bm R}{\partial \alpha}
\right|_*
\delta \alpha=0.
\label{eq:linearized}
\end{equation}
with the comoving linear operator
\begin{equation}
\mathcal{L}\,\delta\bm U
\equiv \mathcal{D}\,\delta\bm U''
+V^*\,\delta\bm U'
+\left.\frac{\partial\bm R}{\partial\bm U}\right|_*
 \delta\bm U ,
\label{eq:Ldef}
\end{equation}
where $\partial\bm R/\partial\bm U|_*$ is the local $2\times 2$ Jacobian of
the reaction terms evaluated along the reference profile~[Eq.~(\ref{eq:Jac})]. Differentiating
Eq.~(\ref{eq:twprofile}) with respect to $z$ shows that
$\mathcal L\bm U_0^{*\prime}=0$: the profile derivative is a right
zero mode, the Goldstone mode of broken translational invariance. Its
existence makes $\mathcal{L}$ singular, so Eq.~(\ref{eq:linearized})
cannot be inverted for $\delta\bm U$; instead a solvability condition
determines $\delta V$.\\

With the inner product
\begin{equation}
\langle\bm P,\bm Q\rangle
\equiv
\int[P_f(z)Q_f(z)+P_g(z)Q_g(z)]\,\mathrm{d}z ,
\label{eq:innerproduct}
\end{equation}
the adjoint operator defined by
$\langle\bm\Psi,\mathcal L\bm\Phi\rangle
=\langle\mathcal L^\dagger\bm\Psi,\bm\Phi\rangle$
follows from integration by parts (boundary terms vanish for a band
localized far from the fixed ends): the diffusion term is
self-adjoint, the advection term changes sign, and the pointwise
Jacobian transposes,
\begin{equation}
\mathcal L^\dagger\bm\Psi
=\mathcal{D}\,\bm\Psi''
-V^*\,\bm\Psi'
+\left(
\left.\frac{\partial\bm R}{\partial\bm U}\right|_*
\right)^T\bm\Psi .
\label{eq:Ldagger}
\end{equation}
The left zero mode satisfies $\mathcal L^\dagger\bm\Psi=0$; the sign
reversal of the advection term is the continuum counterpart of the
reversed-index recursion of the discrete adjoint in
Sec.~\ref{sec:numerics}. Projecting Eq.~(\ref{eq:linearized}) onto
$\bm \Psi$ and using
$\langle\bm\Psi,\mathcal L\,\delta\bm U\rangle
=\langle\mathcal L^\dagger\bm\Psi,\delta\bm U\rangle=0$, the unknown
profile correction drops out, leaving
\begin{equation}
\langle\bm\Psi,\bm U_0^{*\prime}\rangle\,\delta V
=
-\left\langle\bm\Psi,
\left.\frac{\partial\bm R}{\partial \alpha}\right|_*
\right\rangle\delta \alpha .
\end{equation}
Dividing by $\delta \alpha$ gives Eq.~(\ref{eq:fredholm}).

\section{Additional numerical details}
\label{app:details}

The components of the complete fixed-stress right-hand side
$\bm{\mathcal{G}}=(\mathcal{G}_f,\mathcal{G}_g)$ used in the discrete adjoint are
\begin{equation}
\mathcal{G}_f=-\theta e^{-g}f^2,
\label{eq:Ff}
\end{equation}
and
\begin{equation}
\mathcal{G}_g=
\frac{D_g}{\eta}\partial_x^2g
+g_\infty-g-fe^{-g}g.
\label{eq:Fg}
\end{equation}
Endpoint rows of the discrete vector field are set to zero. The transpose Jacobian action combines Eq.~(\ref{eq:Jac}) with the symmetric second-difference operator in $g$. This avoids constructing a dense Jacobian and makes one adjoint solve comparable in cost to one fixed-stress forward trajectory.

The history simulation is examined every ten recorded states. The (nine) strain bins cover $0\le\overline\ep<0.045$. Late-history bins contain slower and less clean bands and consequently show larger scatter in $\mus$; no attempt is made to continue the analysis after the type-A extraction criteria cease to be satisfied. The rate sweep is intentionally restricted to the four values listed in Table~\ref{tab:rate}, because at higher imposed rates the simple peak-based single-band selection becomes increasingly contaminated by irregular or multiple activity.

For the disorder calculation, the first-passage deposited strain on each block is
\begin{equation}
 \Delta\epsilon_i=\Omega\int a_i(\tau)\,\mathrm{d}\tau.
\end{equation}
The ensemble-mean profile is subtracted before forming the cumulative force. Mean-square increments are averaged over disorder realizations and all allowed starting positions. The 10--25~mm fitting interval is at least an order of magnitude larger than the imposed one-block correlation length. The use of the early-history adjoint mobility rather than an independently fitted mobility reduces the disorder calculation to the same collective-coordinate parameters used in the deterministic mapping.

Each disorder realization is integrated to $\tau=180$ with $\Delta \tau = 0.05$. The mean-square force increments are fitted as a linear function of separation over 10--25~mm with the intercept constrained to zero, and $D_{\rm eff}$ is one half of the fitted slope according to Eq.~(\ref{eq:BrownianEmergence}).

%\bibliography{refs}

\end{document}